\documentclass[twocolumn,tighten]{aastex631}

\usepackage{float}
\usepackage{amsmath}	%
\usepackage{todonotes}

\newcommand*\rhocl{\ensuremath{\rho_{\rm cl}}}
\newcommand*\rhow{\ensuremath{\rho_{\rm w}}}
\newcommand*\rhomix{\ensuremath{\rho_{\rm mix}}}

\newcommand*\nw{\ensuremath{n_{\rm w}}}
\newcommand*\emix{\ensuremath{e_{\rm mix}}}
\newcommand*\ecl{\ensuremath{e_{\rm cl}}}
\newcommand*\ew{\ensuremath{e_{\rm w}}}

\newcommand*\emincool{\ensuremath{e_{{\rm min},{\rm cool}}}}
\newcommand*\Tmix{\ensuremath{T_{\rm mix}}}
\newcommand*\Tcl{\ensuremath{T_{\rm cl}}}
\newcommand*\Tw{\ensuremath{T_{\rm w}}}

\newcommand*\mumix{\ensuremath{\mu_{\rm mix}}}
\newcommand*\mucl{\ensuremath{\mu_{\rm cl}}}
\newcommand*\muw{\ensuremath{\mu_{\rm w}}}
\newcommand*\rcl{\ensuremath{R_{\rm cl}}}
\newcommand*\rclcrit{\ensuremath{R_{\rm crit}}}
\newcommand*\rclcritgo{\ensuremath{R_{\rm crit,GO}}}
\newcommand*\rclcritgof{\ensuremath{R_{\rm crit,GOF}}}
\newcommand*\rclcritls{\ensuremath{R_{\rm crit,LS}}}
\newcommand*\vw{\ensuremath{v_{\rm w}}}

\newcommand*\Mw{\ensuremath{\mathcal{M}_{\rm w}}}

\newcommand*\minmix{\ensuremath{{\rm minmix}}}

\newcommand*\tshear{\ensuremath{t_{\rm shear}}}
\newcommand*\tcc{\ensuremath{t_{\rm cc}}}

\newcommand*\tcool{\ensuremath{t_{\rm cool}}}
\newcommand*\tcoolmix{\ensuremath{t_{\rm cool, mix}}}
\newcommand*\tcoolminmix{\ensuremath{t_{{\rm cool},\minmix}}}

\newcommand*\tcoolw{\ensuremath{t_{\rm cool, w}}}

\newcommand*\cscold{\ensuremath{c_{s,{\rm cold}}}}

\newcommand*\cshot{\ensuremath{c_{s,{\rm hot}}}}

\newcommand*\taucool{\ensuremath{\tau_{\rm cool}}}

\newcommand*\taudyn{\ensuremath{\tau_{\rm dyn}}}

\newcommand*\tcoolmin{\ensuremath{t_{\rm cool, min}}}

\newcommand*\LS{``Li/Sparre''}
\newcommand*\GOF{``Gronke/Oh/Farber''}

\newcommand*\enzoe{{\sc enzo-e}}
\newcommand*\enzo{{\sc enzo}}
\newcommand*\cello{{\sc cello}}
\newcommand*\grackle{{\sc grackle}}

\shortauthors{Abruzzo et al.}
\graphicspath{{./}{figures/}}

\begin{document}

\title{TuRMoiL of Survival: A Unified Survival Criterion for Cloud-Wind Interactions}

\author[0000-0002-7918-3086]{Matthew W. Abruzzo}
\affiliation{Department of Astronomy,
Columbia University, 550 West 120th Street,
New York, NY 10027, USA}

\author[0000-0003-3806-8548]{Drummond B. Fielding}
\affiliation{Center for Computational Astrophysics,
Flatiron Institute, 162 5th Avenue,
New York, NY 10010, USA}

\author[0000-0003-2630-9228]{Greg L. Bryan}
\affiliation{Department of Astronomy,
Columbia University, 550 West 120th Street,
New York, NY 10027, USA}
\affiliation{Center for Computational Astrophysics,
Flatiron Institute, 162 5th Avenue,
New York, NY 10010, USA}

\begin{abstract}
Cloud-wind interactions play an important role in long-lived multiphase flows in many astrophysical contexts. When this interaction is primarily mediated by hydrodynamics and radiative cooling, the survival of clouds can be phrased in terms of the comparison between a timescale that dictates the evolution of the cloud-wind interaction, (the {\it dynamical time-scale} $\taudyn$) and the relevant {\it cooling timescale} $\taucool$.
Previously proposed survival criteria, which can disagree by large factors about the size of the smallest surviving clouds, differ in both their choice of \taucool\ and (to a lesser extent) \taudyn. 
Here we present a new criterion which agrees with a previously proposed empirical formulae but is based on simple physical principles. 
The key insight is that clouds can grow if they are able to mix and cool gas from the hot wind faster than it advects by the cloud. 
Whereas prior criteria associate \taudyn\ with the cloud crushing timescale, our new criterion links it to the characteristic cloud-crossing timescale of a hot-phase fluid element, making it more physically consistent with shear-layer studies.
We develop this insight into a predictive expression and validate it with hydrodynamic \enzoe\ simulations of ${\sim}10^4\, {\rm K}$, pressure-confined clouds in hot supersonic winds, exploring, in particular, high wind/cloud density contrasts, where disagreements are most pronounced.
Finally, we illustrate how discrepancies among previous criteria primarily emerged due to different choices of simulation conditions and cooling properties, and discuss how they can be reconciled.
\end{abstract}


\section{Introduction} \label{sec:intro}

A variety of long-lived multiphase flows appearing in galaxy-related contexts involve or derive from cloud-wind interactions, in which a body of cooler gas (\textit{the cloud}) moves with respect to a hotter volume-filling flow (\textit{the wind)}.
Examples of such flows that shape galaxy evolution include multiphase galactic outflows \citep[e.g.,][]{steinwandel22a,rey23a}, the intermediate and high velocity clouds in a galactic fountain \citep[e.g.,][]{tan23a,jennings23a}, or the filamentary accretion of cold intergalactic material \citep[e.g.,][]{mandelker20a}.
Other flows provide valuable information about their progenitor, such as the Magellanic Stream \citep[e.g.,][]{bustard22a} or the tails of jellyfish galaxies \citep[e.g.,][]{tonnesen21a}.

In the absence of cooling the cloud-wind velocity differential drives turbulent mixing that destroys the cloud (i.e. homogenizes it with the other phase) faster than ram pressure removes the differential by accelerating and entraining the cloud. This is fundamentally at odds with the observed long lifetimes of the cool phase of galactic winds \citep[e.g.][]{zhang17a}. However, cooling is observed to be present in many cloud-wind interactions and can allow clouds to survive for long times \citep[e.g.][]{gronke18a}. The key question is what sets the criteria for when cooling outweighs the disruptive effects of mixing, enabling cloud survival?

Cloud survival has received significant attention in the context of galactic winds, in which hot supersonic winds produced by supernovae are expected to accelerate and entrain cool interstellar clouds.
Previous work has explored various physical mechanisms that extend the cloud's lifetime long enough for it to be entrained \citep[e.g.,][]{mccourt15a,gronnow18a,cottle20a} or alter the acceleration mechanism \citep[e.g.,][]{zhang18a,wiener19a,bruggen20a}.

In this letter we consider the regime in which rapid cooling facilitates the long-term survival of clouds \citep[e.g.][]{marinacci10a,armillotta16a}.
We refer to this process as turbulent radiative mixing layer (TRML; which we pronounce as ``turmoil'') entrainment.
This process transfers mass and momentum from the wind to the cloud, which ultimately leads to cloud growth \citep{gronke18a}.
While the precise details have been a topic of vigorous debate, clouds undergo TRML entrainment when they incorporate new hot gas at a rate faster than they lose cold gas. This is generally phrased as the competition between some timescale that dictates the evolution of the cloud-wind interaction, which we generically term the {\it dynamical time-scale}, \taudyn\ with the relevant {\it cooling timescale} \taucool.
Most wind-tunnel studies \citep[e.g.][]{gronke18a,gronke20a,li20a,sparre20a,kanjilal20a,farber22a,abruzzo22a} have adopted for \taudyn\ the timescale on which the cloud is dispersed in the absence of cooling, which is associated with the cloud crushing timescale, although we note that studies of individual shear layers \citep[e.g.,][]{fielding20a,tan21a,chen22a} typically find that \taudyn\ is more closely related to the shear time, which has a different dependence on the cloud/wind density contrast. There has also been significant disagreement over the appropriate definition for \taucool\ largely because the cooling time is phase-dependent and it is not obvious which phase to use when computing the cooling time.

Here we propose a new survival criterion that aims to reproduce the accuracy of previous empirically calibrated criteria with a simple physically motivated argument, thereby resolving many of the disagreements from prior works. In \autoref{sec:methods}, we describe the physical conditions under which we compare survival criteria;
we present the various criteria in \autoref{sec:describe_criteria} and compare them in \autoref{sec:result}.
Finally, we summarize and discuss our results in \autoref{sec:conclusion}.

\section{Scope of this work} \label{sec:methods}

\begin{figure}
  \center
\includegraphics[width = 3.3in]{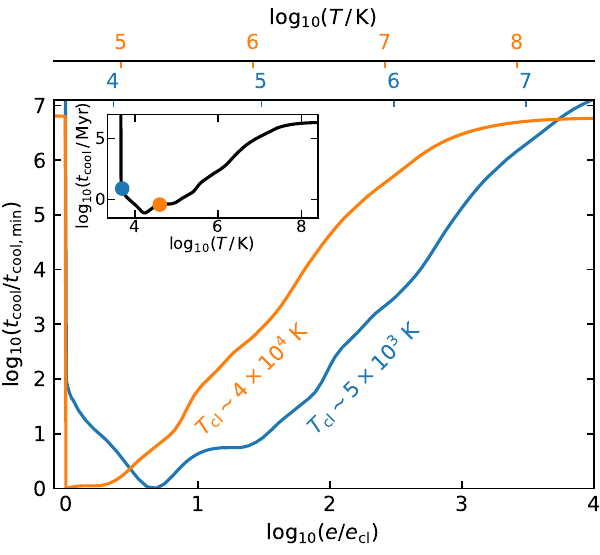}
\caption{\label{fig:tcool}
Illustrates the difference in the shape of $\tcool(e)$ when cooling is turned off below \ecl, for use in our runs with $\Tcl\approx 5\times 10^3\, {\rm K}$ and $\Tcl\approx 4\times 10^4\, {\rm K}$.
The inset shows $\tcool(T)$ as a function of $T$, before any floors are applied, and denotes both \Tcl\ locations with the blue and orange dots.
}
\end{figure}

To maximize our results' generality, %
we focus on TRML entrainment of a pressure-confined cloud in idealized conditions. %
We omit all physics beyond non-dissipative hydrodynamics and optically-thin radiative cooling (at uniform solar metallicity).
While a cooling wind can affect an interaction's dynamics \citep[pressure confinement fails in extreme cases, see][]{li20a}, we consider it a distinct effect from TRML entrainment and take steps to omit it (see below).
We describe regimes where omitted effects are relevant in \autoref{sec:conclusion}.

We restrict this work to supersonic winds ($\Mw=\vw/\cshot>1$), density contrasts ($\chi=\rhocl/\rhow$) of ${\geq}100$, and a uniform initial thermal pressure of $p/k_B = 10^3\, {\rm K}\, {\rm cm}^{-3}$.
We primarily consider clouds with a physically motivated initial temperature of $\Tcl\approx5\times10^3\, {\rm K}$, which is approximately equal to the thermal equilibrium temperature for gas at this pressure exposed to the $z=0$ metagalactic UV background. 
We also consider several cases with $\Tcl\approx4\times10^4\, {\rm K}$ for comparisons with \citet{gronke18a}.
In both cases, we shut cooling off below \Tcl; the modified curves are shown in \autoref{fig:tcool}.
This cooling floor has minimal impact on $\tcool(T)$ for the $\Tcl\approx5\times10^3\, {\rm K}$ case.

All simulations have a resolution of 16 cells per cloud radius and use identical setups and numerical schemes to \citet{abruzzo22b}.
They were run with \enzoe\footnote{\url{http://enzo-e.readthedocs.io}}, a rewrite of \enzo\ \citep{bryan14a} built on the \cello\ AMR framework \citep{bordner12a,bordner18a}.
We modeled radiative cooling with the pre-tabulated solver from the \grackle\ library\footnote{\url{https://grackle.readthedocs.io/}} \citep{smith17a} and assume a $z=0$ \citet{haardt12a} UV background.
We present $\Tcl\approx5\times10^3\, {\rm K}$ simulations taken directly from \citet{abruzzo22b}, supplemented by additional $\chi=100$ runs with $\Mw=3$ and $\Mw=6$.
All $\Tcl\approx4\times10^4\, {\rm K}$ simulations were run for this work.
To avoid cooling of the wind, cooling is artificially turned off above ${\sim}0.6\Tw$ in all cases other than the runs with $\chi=10^4,\Tcl\approx5\times10^3\, {\rm K}$ and $\chi=10^3,\Tcl\approx4\times10^4\, {\rm K}$.\footnote{
These are a subset of our runs where \tcoolw\ is large compared to \tcc. 
More generally, previous works illustrate that turning cooling of the wind off in such cases has negligible impact on cloud survival \citep{gronke18a,abruzzo22a}.
}

\section{Survival Criteria} \label{sec:describe_criteria}

\begin{figure}
  \center
\includegraphics[width = \columnwidth]{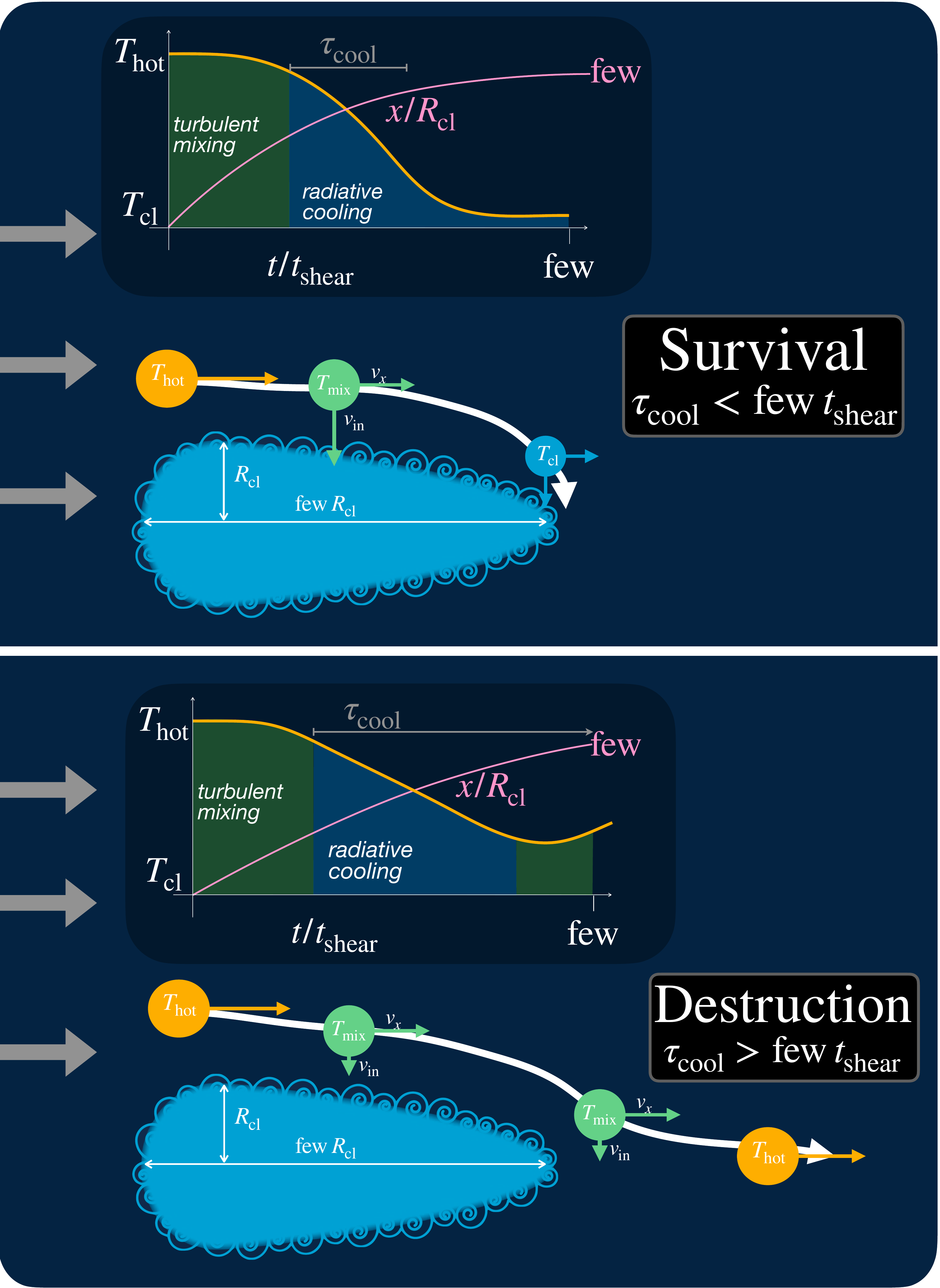}
\caption{\label{fig:cartoon}
A graphical depiction of our proposed new criterion; as discussed in the text, a hot fluid element starts to mix into the cloud, and then must cool before it is advected past the cloud (in a few shear times) in order to be added to the cloud.
}
\end{figure}

\subsection{New Criterion} \label{sec:new_criteria}

We now introduce the physical picture underlying our cloud survival criterion. 
To begin with, there is a large imbalance in the specific thermal energy \citep[or entropy;][]{abruzzo22a} and momentum content of the cloud and wind material.
Turbulent mixing redistributes these imbalances between interacting fluid elements.
Thus, as hot phase fluid elements travel along the length of the cloud, they mix with cooler fluid elements.
This mixing produces intermediate-temperature gas at the interface of the phases, which is highly susceptible to radiative cooling.

The key physical insight is that clouds will not grow (and therefore will eventually be destroyed) if an incoming parcel of hot gas is not able to mix and cool before it is advected past the cloud. In other words,  if the characteristic cooling timescale, \taucool, is short enough then mixed gas will be added to the cool-phase enabling the cloud to grow and ultimately survive.
The window of opportunity for a cloud to capture a fluid element from the wind closes once the fluid element reaches the end of the cloud.
The cartoon in \autoref{fig:cartoon} illustrates that this corresponds to a timescale on the order of (but slightly longer than) $\tshear$, where $\tshear=\rcl/\vw$.
The cloud, therefore, survives when
\begin{equation}
    \label{eqn:criterion_v1}
    \alpha\, \tshear \ga \taucool
\end{equation}
where $\alpha$ is a constant on the order of a few to ${\sim} 10$ that accounts for the fact that the relative velocity is somewhat less than $\vw$ due to momentum mixing, and that the length the element can traverse is somewhat longer than $\rcl$.

We emphasize that this choice for \taudyn\ differs from the usual adoption of the cloud-crushing timescale, which remains the time required for the cloud to be completely destroyed; however, the key point is that clouds need to incorporate fresh hot gas in order to grow, and this process -- and its timescale -- differs from the process of overall cloud destruction and its timescale.

Next, we shift focus to the cooling timescale \taucool, which likely involves a weighted average over the cooling function \citep[e.g.][]{abruzzo22a}.
Like prior works, we approximate it with $\tcool$ computed at a representative intermediate temperature or specific internal energy, $e = p / ((\gamma-1) \rho)$.\footnote{
Throughout this letter, we frame our discussion in terms of specific internal energy, rather than temperature, $T$.
This simplifies equations when the mean molecular weight varies with $T$; for a varying $\mu$, $\Tw<\chi\Tcl$ while $\ew=\chi\ecl$.
The reader can replace all occurrences of $e$ with $T/\mu$ multiplied by some constants.
} 

\citet{gronke18a} make a compelling argument for using the cooling timescale of the gas in the mixing layer $\tcoolmix = \tcool(\emix,p)$.
The definition of \tcoolmix\ derives from the cloud-wind interaction's quasi-isobaric nature and an estimate for the specific internal energy of the mixed gas, $\emix = \sqrt{\ecl\ew}$, motivated by analytic arguments from \citet{begelman90a}. 
While we certainly agree with this overall philosophy, we argue that \tcoolmix\ defined in this way does not always capture the sensitivity of \taucool\ to alterations in the $e$-dependence of $\tcool$ \citep{abruzzo22a} -- in particular, we note that both the gas distribution across phases and the appropriate phase-weighting must depend on the precise form of the cooling function.
A fully self-consistent weighting must await future work so here we adopt a subtle but important refinement to this original estimate in which $\taucool\approx\tcoolminmix=\tcool(\sqrt{\chi \emincool \ecl},p)$ \citep{farber22a}, where \emincool\ coincides with the minimum $\tcool$ for $\ecl\leq e\leq\ew$.\footnote{For the cooling function used in our $\Tcl\approx 4\times10^4\, {\rm K}$ runs (as was used in the original \citet{gronke18a} paper), $\tcoolminmix = \tcoolmix$ since \emincool, the $e$ that coincides with the minimum  $\tcool$ over $\ecl\leq e\leq\ew$, matches \ecl.}

Now, we plug our choice for $\taucool$ into \autoref{eqn:criterion_v1}.
Our new criterion predicts cloud survival for
\begin{equation}
    \label{eqn:tcomp_new}
    \alpha\, \tshear \ga \tcoolminmix =  \tcool(\sqrt{\chi\emincool\ecl}, p).
\end{equation}
We find that $\alpha\sim7$ provides a good fit to the data and is consistent with our physical picture.
This predicts that clouds survive when their radius exceeds the critical value 
\begin{equation}
    \label{eqn:rcrit_new}
    \rclcrit = \cscold\, \Mw\, \sqrt{\chi}\, \tcool(\sqrt{\chi\emincool\ecl}, p)/\alpha,
\end{equation}
where $\cscold=\sqrt{\gamma(\gamma - 1)\ecl}$ is the initial sound speed within the cloud.

\subsection{Existing Criteria}

To facilitate comparisons, we now briefly review some previously proposed survival criteria.
The existing criteria are conceptually different from our new proposed criterion. 
As opposed to comparing cooling to advection, they argue that a cloud survives when cooling is more rapid than the rate at which it is destroyed. 
The cloud destruction timescale, is generally linked to $\tcc=\sqrt{\chi}\tshear$.
Realistic galactic environments commonly have $10^2 \la \chi \la 10^4$.

\citet{gronke18a} proposed the first survival criterion, $\tcc > \tcoolmix$, and \citet{farber22a} later proposed an improved refinement, $\tcc>\tcoolminmix$, that we term the \GOF\ criterion.
These criteria specify critical cloud radii of 
\begin{eqnarray}
    \rclcritgo  & \sim & \cscold\, \Mw\, \tcool(\sqrt{\chi}\ecl, p)           \label{eqn:rcrit_go}\\
    \rclcritgof & \sim & \cscold\, \Mw\, \tcool(\sqrt{\chi\emincool\ecl}, p). \label{eqn:rcrit_gof}
\end{eqnarray}

\citet{li20a} proposed a separate criterion $f\tcc>\tcoolw=\tcool(\ew,p)$, where $f(\nw, R_{\rm cl}, \vw)$ is an empirical parameterization of cloud lifetime described by
\begin{equation}
    \left(\frac{f}{9\pm1}\right)^{10/3} = \eta\,  
    \frac{2 \rcl}{1\, {\rm pc}}\ \frac{\nw}{10^{-2}\, {\rm cm}^{-3}}\ \frac{\vw^2}{10^4\, {\rm km}^2\, {\rm s}^{-2}},
\end{equation}
when $\eta=1$.
We actually focus on a variant proposed by \citet{sparre20a} that improves accuracy when $\Mw>1$, by using $\eta = 2 (\Mw/1.5)^{-2.5}$.
This \LS\ criterion, gives a critical cloud radius of
\begin{eqnarray}
    \label{eqn:rcrit_ls}
    \rclcritls = 
        72.2\, {\rm pc}\,
        \left(\frac{\gamma}{5/3}\ \frac{0.6}{\mu_{\rm w}}\ \frac{p/k_B}{10^3\, {\rm K}\, {\rm cm}^{-3}}\right)^{-3/13}
        \nonumber \\
        \left(\Mw^{2.9} \frac{\tcool(\chi \ecl, p)}{{\rm Gyr}}  \frac{\cscold}{10\, {\rm km}\, {\rm s}^{-1}}\right)^{10/13} 
    .
\end{eqnarray}

\section{Results and Reconcillation} \label{sec:result}

\begin{figure}
  \center
\includegraphics[width = 3.3in]{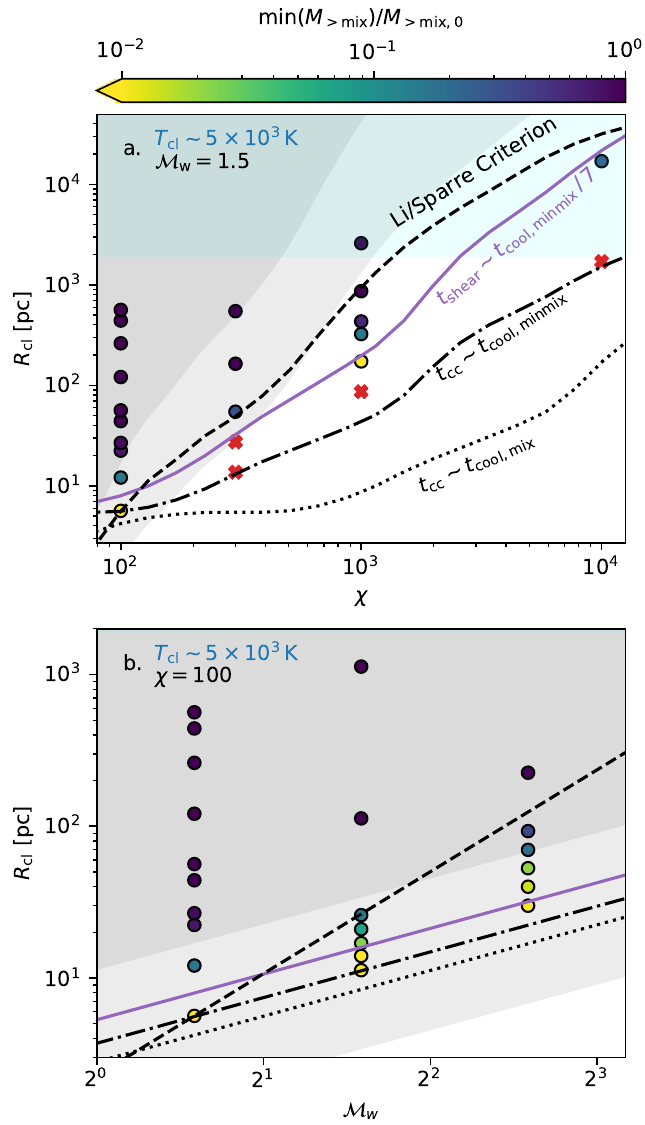}
\caption{\label{fig:survival-criteria-5e3}
Illustrates the fates of our $\Tcl\approx5\times 10^3\, {\rm K}$ cloud-wind simulations.
The circles are colored by the minimum cool gas mass (i.e. gas with $\rho>\rhomix=\sqrt{\chi}\rhocl$) in a simulation (before it starts growing) while the red ``X''s show runs where this mass drops to 0.
Runs in panel a (b) have $\Mw=1.5$ ($\chi =100)$ and vary the value of $\chi$ ($\Mw$).
The black dotted (dashed-dotted) line shows the \citet{gronke18a} survival criterion with (without) the \citet{farber22a} correction.
The black dashed line shows the \citet{li20a} criterion with the \citet{sparre20a} correction.
The magenta line shows our preferred survival criterion.
While all simulations are idealized, shaded regions show where other effects become important in more realistic simulations.
A cooling wind may be dynamically relevant in the light gray region ($\tcoolw/\tcc <100$).
In the dark gray region ($\tcoolw < \rcl/\max(\vw,\cscold)$) or cyan region (\rcl\ exceeds the Jeans length, $\lambda_J=\cscold/\sqrt{G\rho}$), a cooling wind or self-gravity dominates the interaction \citep{li20a}.
}
\end{figure}

\begin{figure}
  \center
\includegraphics[width = 3.3in]{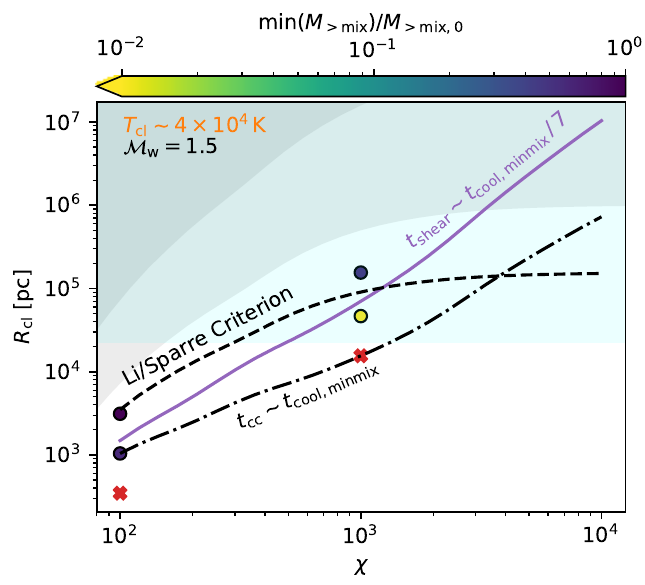}
\caption{\label{fig:survival-criteria-4e4}
Similar to \autoref{fig:survival-criteria-5e3}a, except that all runs have $\Tcl\approx4\times 10^4\, {\rm K}$.
In this figure, the $\tcc=\tcoolminmix$ and $\tcc=\tcoolmix$ curves are identical.
}
\end{figure}

We now turn to the key question: how well does this criterion work in simulations over a wide parameter range?
Fig. \ref{fig:survival-criteria-5e3} and \ref{fig:survival-criteria-4e4} illustrate the robustness of our new survival criterion, \autoref{eqn:tcomp_new}. Simulations with various parameter choices are shown as symbols coded by their survability, while the lines show the predicted minimum cloud radius for survival.
In these plots, rather than a binary categorization of the clouds' fates, we show the minimum cool phase mass, before the cloud starts growing.\footnote{There are two cases where the cool phase mass remains positive, but never show growth. 
However, in both cases the minimum mass is under $1\%$ of the initial value.}
As in prior works \citep[e.g.][]{li20a,sparre20a}, we define the cool phase as all gas with $\rho>\rhomix = \rhocl/\sqrt{\chi}$. %
We consider clouds to be destroyed when the cool-phase mass drops below some threshold fraction of its initial value.
Our new criterion clearly matches the simulation results for sensible choices of the threshold (i.e., values ranging from $1\%$ to $10\%$). In particular, it reproduces the steep $\chi$ dependence on $R_{\rm crit}$ that is seen in the numerical simulations and is also consistent with the results when varying Mach number.

One natural question is how this criterion compares to those that have been previously suggested and how to reconcile any differences. One possibility is that the interpretation of the simulation results might differ.
Indeed, the exact definition of cloud destruction in a simulation is a thorny topic\footnote{For example, it may be important to differentiate between the case where a cloud survives and the case where cloud material is converted to hotter gas that seeds prompt cool-phase precipitation at a downstream location.
This distinction has important implications for the transport of dust in galactic winds.
However, such an endeavor is well beyond the scope of this letter.}
without an obvious definition; there are almost as many definitions as there are survival criteria.
These differing definitions have sometimes been cited as potential explanations for discrepant predictions made by survival criteria \citep[e.g.][]{kanjilal20a,sparre20a,farber22a}; however, as we will demonstrate below, we generally find agreement when the criteria are appropriately compared, and these definitional differences are of secondary importance.

To understand the relationships between survival criteria, it is insightful to compare the dependence on parameters describing the cloud-wind interaction, rather than the exact normalization.
For interactions with $\Tcl\approx5\times10^3\, {\rm K}$ and $100\leq\chi\leq10^4$,
we find\footnote{Under these conditions, \muw\ is constant and $\tcool\propto e^{2.4} p^{-1}$ for %
the relevant values of $\sqrt{\emincool \ew}$ and $\ew$.} 
\begin{eqnarray}
\rclcritgof  & \propto &  \chi^{1.2}\, \Mw\, p^{-1} \\
\label{eqn:ls_scaling}
\rclcritls   & \propto &  \chi^{1.85}\, \Mw^{2.23}\, p^{-1} \\
\label{eqn:new_scaling}
R_{\rm crit,new}  & \propto &  \chi^{1.7}\, \Mw\, p^{-1},
\end{eqnarray}
where $R_{\rm crit,new}$ corresponds to our new scaling relation, \autoref{eqn:rcrit_new}.
Comparisons of $\chi$ dependence are of central importance.

We start by comparing the $\chi$ dependence of the various criteria to the results of our simulations. 
\autoref{fig:survival-criteria-5e3}a and \ref{fig:survival-criteria-4e4} illustrate that the steeper $\chi$ dependence of our new criterion and the \LS\ criterion both better match the observed high-$\chi$ cloud survival than the \GOF\ criterion.
However, although both criteria predict the same $\chi$ scaling, the physical picture behind the strong $\chi$ scaling in our criterion is quite different from that of the \LS\ criterion. 
The \LS\ criterion essentially argues that \taucool\ should heavily weight high-$T$ cooling (i.e. cooling in the hot wind itself).
While $\taucool\approx\tcoolmix$ may over-emphasize the low-$T$ cooling in some cases, given the fact that many of our simulations shut off high-$T$ cooling by hand and still find the same scaling, it is clear that $\taucool\approx\tcoolw$ incorrectly highlights the importance of cooling at high-$T$. Interestingly, since it matches the simulation results, we suggest that  the accurate predictions of the \LS\ criterion arise because of a surprising cancellation: the adoption of $\taucool\approx\tcoolw$ compensates for the incorrect $\chi$-dependence of $\taudyn\sim\tcc$.

We now compare the results of simulations to the various predictions for the \Mw\ scaling of the critical cloud radius. 
\autoref{fig:survival-criteria-5e3}b illustrates that there is some dependence on the definition of cloud destruction, with more relaxed definitions (i.e. allowing small survival fractions to count as survived) favoring our criterion's relatively shallow scaling, and stricter definitions (i.e. requiring higher survival fractions) favoring an intermediate dependence.
While the \LS\ criterion favors a steeper \Mw\ dependence, the supersonic simulations from \citet{sparre20a} are consistent with our result.
The consistency of their $\Mw\geq1.5$ runs with an intermediate \Mw\ dependence becomes apparent when the clouds' fates, specified by their fairly strict definition of cloud destruction, are plotted as a function of \Mw\ and \rcl.
Their simulations only favor a steeper \Mw\ dependence when their $\Mw=0.5$ runs are also considered.
This is not entirely surprising given the evidence that the interaction changes for subsonic winds \citep[e.g.][]{scannapieco15a,yang23a}.
We expect followup work to show that \rclcrit\ has steeper \Mw\ scaling for subsonic winds.

Finally, we explain how criteria with weaker $\chi$-dependence emerged.
\citet{gronke18a} focused on simulations with $\Tcl\approx4\times 10^4\, {\rm K}$ and $100 \leq \chi \leq 1000$, and, as illustrated in \autoref{fig:survival-criteria-4e4}, reached the very reasonable conclusion that $\tcoolmix<\tcc$ describes cloud survival.
At this \Tcl\ (essentially at the peak of the cooling curve), $\tcoolmin=\tcoolminmix$ and the only difference with our criterion is the use of \tcc\ in place of \tshear.
Due to the exploratory nature of their work, \citet{gronke18a} varied cloud radius by factors of ${\approx}10$ at each considered $\chi$; while this allowed them to sample a large region of parameter space, it was too sparsely sampled to capture the $\sqrt{\chi}$ difference in scaling between \tcc\ and \tshear.
When \citet{farber22a} determined that \tcoolminmix\ should replace \tcoolmix, they used runs that varied \Tcl\ for fixed $\chi$. %
Because $\chi$ was fixed, $\taudyn=\tshear$ and $\taudyn=\tcc$ are equally consistent with their results.

There have been several works that highlighted differences between the \citet{gronke18a} and \LS\ criteria \citep[e.g.][]{sparre20a,kanjilal20a}, however these came before the $\tcoolminmix$ refinement introduced by \citet{farber22a}.
Reconciliation efforts were stymied by the fact that only the combined differences in \taudyn\ and \taucool\ are sufficient to explain the \citet{gronke18a} criterion's weaker $\chi$-dependence for $\Tcl\approx10^4\, {\rm K}$ (i.e. the commonly considered \Tcl).
Further challenges likely arose from the effectiveness of $\taucool\sim\tcoolmix$ for $\Tcl\approx 4\times 10^4\, {\rm K}$ \citep[considered by][]{gronke18a} and association of \taudyn\ with \tcc\ by both criteria.

\section{Conclusion} \label{sec:conclusion}

In this letter we propose a new criterion for predicting the survival of cool (${\sim}10^4\, {\rm K}$) clouds that encounter hot supersonic flows, based on the ratio between the timescale for hot-phase fluid elements to cross the length of the cloud, $\alpha\tshear$, and \taucool, an estimate for the weighted average of the cooling timescale.
Our new criterion shares the accuracy of the more empirically calibrated criteria from \citet{li20a} and \citet{sparre20a}, particularly in the high-$\chi$ limit, while employing the better-physically motivated choice of $\taucool\sim\tcoolminmix$, as in the \citet{gronke18a,farber22a} criterion.
Additionally, discrepancies among criteria {\it primarily} emerged/persisted due to choices of validation simulation conditions ($\Tcl,\rcl,\chi$), while differences in the definition of cloud destruction within the simulations turn out to be less important.

While previous survival criteria compare \taucool\ against \tcc, we are encouraged that our choice to compare against $\alpha\tshear$ produces more consistent scaling compared to the eddy turnover timescale, which is the relevant dynamical timescale on the micro-scale \citep[e.g.][]{fielding20a, tan21a, chen22a}.
We are also encouraged that tall-box simulations of more realistic cloud-wind interactions find cloud-size distributions consistent with our new criterion \citep{tan23b}.

We predict cloud survival when $\alpha\tshear\ga \tcoolminmix$, for $\alpha\sim7$.
We tested our criterion for $5\times10^3\lesssim (\Tcl/{\rm K}) \lesssim 4\times 10^4$ and supersonic winds, under the assumption that cooling in the wind is unimportant.
Our criterion provides a minimum survival radius of
\begin{eqnarray}
  \rclcrit &=& 271\, {\rm pc} \frac{\mathcal{M}_{\rm w} \chi_3^{1.5} c_{{\rm s,cold},1}^3}
          {\hat{p}_3 \Lambda_{\minmix,-21.4}}\,
          \frac{T_{\rm min,cool}}{10^{4.25} {\rm K} }\,
          \frac{0.71}{\mu(T_{\rm min,cool},p) }\,
          \frac{7}{\alpha} \nonumber \\
  \label{eqn:rcrit_numbers-approx}
  &\approx& 618\, {\rm pc} \frac{\mathcal{M}_{\rm w} \chi_3^{1.5}\, T_{{\rm cl},4}^{1.5}}
      {\hat{p}_3 \Lambda_{\minmix,-21.4}}\,
     \frac{T_{\rm min,cool}}{10^{4.25}\, {\rm K}},
\end{eqnarray}
where $\Lambda_{\minmix,-21.4} \equiv \Lambda(T_\minmix)/(10^{-21.4}\, {\rm erg}\, {\rm cm}^3\, {\rm s}^{-1})$, $\hat{p}_3\equiv p k_B^{-1}/(10^3 {\rm K}\, {\rm cm^{-3}})$, $\chi_3\equiv\chi/10^3$, $c_{{\rm s,cold},1} \equiv \cscold/(10\, {\rm km}\, {\rm s}^{-1})$, $T_{\rm cl,4}\equiv \Tcl/(10^4\, {\rm K})$, and $T_{\rm min,cool}$ is the temperature where \tcool\ is minimized (between \Tcl\ and \Tw).
In calculations where the temperature dependence of $\mu$ is modeled, the first line of \autoref{eqn:rcrit_numbers-approx} is accurate, while approximations in the second line reduce accuracy to within ${\approx}30\%$ for $\Tcl\ga 5\times10^3\, {\rm K}$ and our fiducial pressure $\hat{p}_3 = 1$.\footnote{
This error primarily arises from the omission of the $\mucl^{-1.5}$ dependence.
Additional error can arise at other pressures from the omitted $\mu_{\rm min,cool}^{-1}$ dependence.
When $\Tcl<10^4\, {\rm K}$, larger errors can arise when the value of $T_\minmix$, used to compute $\Lambda(T_\minmix)$, is estimated while ignoring $\mu$ dependence.
}

Our criterion's predictions are most meaningful for clouds in thermal equilibrium (for reasonable equilibration timescales).
For our fiducial pressure, $\hat{p}_3 = 1$, $T_{\rm min,cool}=10^{4.25}\, {\rm K}$, the equilibrium \Tcl\ is ${\approx}5\times 10^3\, {\rm K}$, and $\cscold\approx 8.5\, {\rm km}\, {\rm s}^{-1}$.
With that said, our criterion can still be applied at other choices of \Tcl\ under the assumption that cooling is shut off below that temperature.\footnote{
    We expect our criterion to be accurate for any spatially uniform $\tcool(T)$ that satisfies three properties on the interval $\Tcl< T< \Tw$: (i) it has no local maximum, (ii) there is up to 1 local minimum, which lies between \Tcl\ and $\Tmix=\sqrt{\chi}\mumix\Tcl/\mucl$, and (iii) cooling in the wind is relatively slow.}
Under such contrived conditions, with $\hat{p}_3 = 1$ and $\Tcl>5\times 10^3\, {\rm K}$, $T_{\rm min,cool}=\max(\Tcl,\, 10^{4.25}\, {\rm K})$.

Given that lower pressure environments have larger equilibrium \Tcl, larger $T_{\rm min,cool}$ and smaller magnitudes of $\Lambda$, our criterion suggests that clouds likely require unreasonably large sizes to survive in such cases.
In contrast, our new survival criteria is likely relevant for colder clouds, like those expected in the higher pressure environments (the equilibrium temperature significantly drops for $p/k_B\gtrsim 10^4\, {\rm K}\, {\rm cm}^{-3}$) where starburst galaxies launch their outflows \citep{fielding22a}.

Based on the recent study of colder (${\la} 10^3\, {\rm K}$) gas clouds in \citet{farber22a}, we expect that our criterion specifies conditions for the long-term survival of a cool (${\sim}10^4\, {\rm K}$) phase and suspect that $\tshear\, \psi \ga t_{\rm cool,max}$ may specify conditions for the colder phase's survival; $\psi$ is a normalization constant and $t_{\rm cool,max}$ is the maximum of \tcool\ between \Tcl\ and $T_{\rm min,cool}$. 
Because \citet{farber22a} considered runs with fixed $\chi$, their results fit these criteria equally as well as their own criteria (the primary difference is they use \tcc\ instead of \tshear).
It is also noteworthy that the \LS\ criterion does not generalize well to these conditions \citep[see Fig. 8 of][]{farber22a}.
Followup work is clearly required.

Next, we highlight limitations of our idealized setup.
The shaded regions in Fig. \ref{fig:survival-criteria-5e3} and \ref{fig:survival-criteria-4e4} illustrate where other effects are relevant in more realistic simulations.
The light gray region's lower bound highlights where a cooling wind may become dynamically important \citetext{e.g. \citealt{abruzzo22a} shows a case where equilibrium pressure drops without affecting cool-phase mass evolution}, and the dark gray region highlights where pressure confinement of the cloud fails \citep{li20a}.
The cyan region highlights Jeans-unstable clouds that will likely collapse \citep{li20a}.
Self-shielding against ionizing radiation may also be dynamically relevant at most radii.

In other words, simulations of cool (${\sim}10^4\, {\rm K}$) clouds that undergo TRML entrainment in a supersonic flow only make robust predictions for a narrow range of conditions, while neglecting self-gravity, cooling in the wind, and self-shielding.
However, these assumptions may be robust for subsonic flows, if the survival criterion has steeper \Mw\ dependence in those cases.
This regime is of particular interest in the context of starburst galaxy outflows because most cloud acceleration occurs when the outflow is subsonic \citep[e.g.][]{fielding22a}.

Future work may also want to consider the impact of other physical effects.
For example, there is evidence that magnetic fields with intermediate to strong field-strengths may make it easier for clouds to survive \citep[e.g.][]{hidalgo-pineda23a}.
While the effects of conduction \citep[e.g.][]{li20a} could have large impacts, there's reason to believe they won't strongly influence the survival criterion \citep[e.g.][]{tan21a}.
There are other context-specific effects that may also be relevant, such as external sources of turbulence \citep[e.g.][]{schneider20a,gronke22a}, large-scale external gravitational fields \citep[e.g.][]{tan23a}, cloud geometry \citep[e.g.][]{mandelker20a,bustard22a}, or ensembles of clouds \citep[e.g.][]{aluzas12a,forbes19a,banda-barragan20a}.

Finally, we consider convergence of the cool phase mass, $M_{> {\rm mix}}$, evolution.
While our resolution, 16 cells per cloud radius, is sufficient for convergence in most cases, convergence is more difficult when $\Mw=6$ \citep{gronke20a} or $\chi=10^4$ \citep[e.g][]{abruzzo22b}.
For $\Mw=6$, we are encouraged by convergence in the $M_{> {\rm mix}}$ evolution (at low resolutions) of the case taken from \citet{abruzzo22b} and the monotonic scaling of the minimum $M_{> {\rm mix}}$ with \rcl\ (see \autoref{fig:survival-criteria-5e3}b). 
Even if the $\chi=10^4$ runs are not converged, they are not of central importance to our conclusions.

\begin{acknowledgments}
We are grateful to R. Farber, M. Gronke, P. Oh, and B. Tan for useful conversations about survival criteria.
We are also thankful for the efforts of J. Bordner,  M. Norman and the efforts of the other \enzoe\ developers. GLB acknowledges support from the NSF (AST-2108470, XSEDE grant MCA06N030), NASA TCAN award 80NSSC21K1053, and the Simons Foundation (grant 822237) and the Simons Collaboration on Learning the Universe. 
\end{acknowledgments}

\vspace{5mm}

\software{numpy \citep{harris20a},
          matplotlib \citep{hunter07a},
          yt \citep{turk11a},
          scipy \citep{virtanen20a},
          pandas \citep{mckinney10a}, %
          }

\bibliography{ref}{}
\bibliographystyle{aasjournal}

\end{document}